\documentclass{elsart1p}
\usepackage{graphicx}
\usepackage{amssymb}
\begin{document}
\begin{frontmatter}
%
% Title, authors and addresses
%
% use the thanksref command within \title, \author or \address for footnotes;
% use the corauthref command within \author for corresponding author
% footnotes;
% use the ead command for the email address,
% and the form \ead[url] for the home page:
% \title{Title\thanksref{label1}}
% \thanks[label1]{}
% \author{Name\corauthref{cor1}\thanksref{label2}}
% \ead{email address}
% \ead[url]{home page}
% \thanks[label2]{}
% \corauth[cor1]{}
% \address{Address\thanksref{label3}}
% \thanks[label3]{}
%
\title{New Method of Modelling Dissipative Hydrodynamics}
%
% use optional labels to link authors explicitly to addresses:
% \author[label1,label2]{}
% \address[label1]{}
% \address[label2]{}
%
\author[1]{T. Osada}
\ead{osada@ph.ns.musashi-tech.ac.jp} and \author[2]{G.
Wilk\thanksref{label2}}\ead{wilk@fuw.edu.pl}
\thanks[label2]{Presenting Author.}
\address[1]{Theoretical Physics Laboratory,
Faculty of Knowledge Engineering, Musashi Institute of Technology,
Setagaya-ku, Tokyo 158-8557, Japan}
\address[2]{The Andrzej So{\l}tan Institute for Nuclear Studies,
Ho\.{z}a 69, 00681, Warsaw, Poland}
\begin{abstract}
We propose to model the dissipative hydrodynamics used in
description of the multiparticle production processes
($d$-hydrodynamics) by a special kind of the perfect nonextensive
fluid ($q$-fluid) where $q$ denotes the nonextensivity parameter
appearing in the nonextensive Tsallis statistics. The advantage of
$q$-hydrodynamics lies in its formal simplicity in comparison to
the $d$-hydrodynamics. We argue that parameter $q$ describes
summarily (at least to some extent) all dynamical effects behind
the viscous behavior of the hadronic fluid.
\end{abstract}
\end{frontmatter}

The ideal hydrodynamic model successfully reproduces most of the
RHIC experimental data \cite{Hirano0808.2684} suggesting that
matter created there resembles a strongly interacting quark-gluon
fluid \cite{TDLeeNPA750} rather than a free parton gas. However,
there are also hints that, after all, a hadronic fluid cannot be
totally ideal \cite{nonviscousindic} and should be described by
some kind of dissipative hydrodynamic model (or $d$-hydrodynamics)
\cite{EHI}. A number of such models have been proposed recently
\cite{dhydro,Problems}. Although very promising, they have many
difficulties, both in what concerns their proper formulation and
applications \cite{Problems}. It turns out, however, that one can
obtain constitutive equations of causal $d$-hydrodynamics starting
from the relativistic perfect nonextensive hydrodynamics
($q$-hydrodynamics) supplemented by the nonextensive/dissipative
correspondence (NexDC) proposed in \cite{OsadaPRC77} and taking
place between the perfect $q$-hydrodynamics (based on the
nonextensive Tsallis statistics \cite{Tsallis} and described by
the nonextensivity parameter $q$) and the usual $d$-hydrodynamics
(based on the extensive Boltzmann-Gibbs statistics (BG) to which
Tsallis statistics converges for $q \rightarrow 1$). .

Our $q$-hydrodynamics is causal and preserves the simplicity of
$d$-hydrodynamics facili\-tating its numerical applications
\cite{OsadaPRC77}. It is based on the relativistic nonextensive
kinetic theory proposed in \cite{LavagnoPhysLettA301}, which
replaces the usual notion of local thermal equilibrium by a kind
of stationary state. This stationary state includes, by means of
the parameter $q$ \cite{Stationary}, all kinds of possible strong
intrinsic fluctuations and long-range correlations existing in a
hadronizing system. In this approach the {\it perfect
$q$-hydrodynamic equations} (i.e., equations for the perfect
nonextensive $q$-fluid without any additional currents) are given
by:
\begin{eqnarray}
{\cal T}_{q;\mu}^{\mu\nu} = \Big[ \varepsilon_q(T_q)
u_q^{\mu}u_q^{\nu} - P_q(T_q) \Delta_q^{\mu\nu} \Big]_{;\mu} = 0,
\label{eq:q-equation_of_motion}
\end{eqnarray}
where ${\cal T}_{q}^{\mu\nu}$ is $q$-version of the energy
momentum tensor with  $\varepsilon_q(T_q)$, $P_q(T_q)$,
$T_q=T_q(x)$ and $u^{\mu}_q(x)$ being, respectively, the
$q$-versions of the nonextensive energy density, pressure and
temperature field \cite{OsadaPRC77} and an accompanying
hydrodynamic $q$-flow four vector field (which allows
decomposition (\ref{eq:q-equation_of_motion}) to be performed;
$\Delta_q^{\mu\nu} \equiv g^{\mu\nu}-u_q^{\mu}u_q^{\nu}$). If
$T(x)$ and $u^{\mu}(x)$ are the corresponding temperature and
velocity fields in the case of BG statistics ($q=1$) with
$\varepsilon$ and $P$ being the usual energy density and pressure
then, as proposed in \cite{OsadaPRC77}, one can map a $q$-flow
(with $q$-temperature $T_q$ \cite{OsadaPRC77}) into some
dissipative $d$-flow by requiring that the following relations
hold:
%%%%%%%%%%%%%%%%%%%%%%%%%%%%%%%%%%%%%%%%%%%%%%%%%%%%%%
\begin{eqnarray}
P(T) = P_q(T_q),\quad \varepsilon(T) = \varepsilon_q(T_q) + 3\Pi,
\label{eq:Nex/diss}
\end{eqnarray}
%%%%%%%%%%%%%%%%%%%%%%%%%%%%%%%%%%%%%%%%%%%%%%%%%%%%%%
where $\Pi$ is to be regarded as the bulk pressure of
$d$-hydrodynamics which is defined by
\begin{eqnarray}
\Pi  \equiv \frac{1}{3} w_q [\gamma^2+2\gamma] .
\label{eq:bulk_pressure}
\end{eqnarray}
Here $\delta u_q^{\mu} \equiv u_q^{\mu}-u^{\mu}$ is the four
velocity difference between the non-extensive parameter $q>1$ and
$q=1$ flows, $\gamma(x) \equiv \delta u_q^{\mu}(x) u_{\mu}(x) =
-\frac{1}{2}\delta u_{q\mu} \delta u_q^{\mu}$ and $w_q\equiv
\varepsilon_q+P_q$ is the $q$-enthalpy.

Now, it can be shown \cite{OsadaPRC77} that, if $T(x)$ and
$\gamma(x)$ satisfy relations (\ref{eq:Nex/diss}) in the whole
space-time region, one can always transform equation of perfect
$q$-hydrodynamics Eq.~(\ref{eq:q-equation_of_motion}) into the
following equation of $d$-hydrodynamics \cite{dhydro}:
\begin{eqnarray}
\left[ \varepsilon(T) u^{\mu}u^{\nu} \!-\!\left[ P(T)+\Pi \right]
\Delta^{\mu\nu} \!\!\!+\! 2 W^{(\mu} u^{\nu )} \!+\!\pi^{\mu\nu}
\right]_{;\mu} \!\!\!\!=0, \label{eq:Nex/diss_equation}
\end{eqnarray}
in which one recognizes the energy flow vector $W^{\mu}$ and the
(symmetric and traceless) shear pressure tensor $\pi^{\mu \nu}$,
\begin{equation}
W^{\mu} = w_q[1+\gamma] \Delta^{\mu}_{\sigma} \delta
u_q^{\sigma},\quad \pi^{\mu\nu} = w_q \delta u_q^{<\mu} \delta
u_q^{\nu >} \label{eq:Wpi}
\end{equation}
(where $\Delta^{\mu\nu} \equiv g^{\mu\nu}-u^{\mu}u^{\nu}$, $A^{(
\mu}B^{\nu )}\equiv \frac{1}{2}(A^{\mu}B^{\nu} +A^{\nu}B^{\mu})$,
and $a^{<\mu}b^{ \nu >} \equiv [\frac{1}{2}(\Delta^{\mu}_{\lambda}
\Delta^{\nu}_{\sigma} +
  \Delta^{\mu}_{\sigma} \Delta^{\nu}_{\lambda} )
  -\frac{1}{3}\Delta^{\mu\nu} \Delta_{\lambda\sigma} ]
  a^{\lambda}b^{\sigma}$). The $d$-hydrodynamics represented
by Eq.~(\ref{eq:Nex/diss_equation}) can be therefore regarded as a
viscous counterpart of the perfect $q$-hydrodynamics represented
by Eq.~(\ref{eq:q-equation_of_motion}).  We call this relation the
{\it nonextensive/dissipative correspondence}, (NexDC). With the
bulk pressure Eq.~(\ref{eq:bulk_pressure}) and NexDC relations
Eq.~(\ref{eq:Nex/diss}) one gets the $q$-enthalpy $w_q =
w/[1+\gamma]^2$ with $w\equiv Ts = \varepsilon+P$ being the usual
enthalpy in BG statistics. Accordingly, the bulk pressure
Eq.~(\ref{eq:bulk_pressure}) can be written as
\begin{eqnarray}
\Pi =  w  \Gamma,\quad {\rm where}\quad \Gamma\equiv
\frac{1}{3}\frac{\gamma(\gamma+2)}{(\gamma + 1)^2}.
\label{eq:Gamma}
\end{eqnarray}
Finally, the NexDC leads to the following relations between
components of the dissipative tensor:
\begin{equation}
W^{\mu} W_{\mu} \!=\! -3\Pi w,~ \pi^{\mu\nu}W_{\nu} \!=\! -2 \Pi
W^{\mu},~  \pi_{\mu\nu}\pi^{\mu\nu} \!=\! 6\Pi^2. \label{eq:pipi}
\end{equation}

Referring to \cite{OsadaPRC77} for details and examples of
numerical applications, we would like to close with two remarks.
First, albeit in our $q$-hydrodynamic the $q$-fluid is ideal
(there is no production of $q$-entropy), the usual entropy in the
corresponding $d$-fluid is produced (as result, the total
multiplicity increases with $(q-1)$). Second, this approach
results in a very simple relation between the ratios of shear and
bulk viscosities over the entropy density, $\eta/s$ and $\xi/s$,
respectively:
\begin{eqnarray}
 \frac{1}{\zeta/s}+\frac{3}{\eta/s} = \frac{w \sigma^{\mu}_{{\rm
 full};\mu}}{\Pi^2}.
\label{eq:sum_rule}
\end{eqnarray}
where $\sigma^{\mu}_{\rm full}$ is the full order dissipative
entropy current. Further investigations of $q$-hydrodynamics and
its implications are under way.

\section*{Acknowledgment}

GW would like to express his gratitude towards the organizers of
the PANIC2008 conference (Eilat, Israel, 10-14 November 2008)
where this work was presented. Partial support (GW) from the
Ministry of Science and Higher Education under contract 1P03B02230
is also acknowledged.

\end{document}